\documentclass[twocolumn,amsmath,showpacs,amssymb,amsfonts,floatfix]{revtex4}

\usepackage{color}
\usepackage{graphicx}
\usepackage{dcolumn}
\usepackage{bm}
\usepackage{soul}
\usepackage{bbm}
\usepackage{ulem}

\usepackage[english]{babel}
\usepackage{amsmath}
\usepackage{gnuplottex} 

\usepackage{graphicx}

\begin{document}
\title{Majorana fermions from Shiba states in an antiferromagnetic chain \\ on top of a superconductor}
\author{Andreas Heimes}
\email{andreas.heimes@kit.edu}
\author{Panagiotis Kotetes}
\author{Gerd Sch\"on}
\affiliation{Institut f\"ur Theoretische Festk\"orperphysik and DFG-Center for Functional Nanostructures (CFN), Karlsruhe Institute of Technology, D-76128 Karlsruhe, Germany}

\begin{abstract} 
We propose a new mechanism for topological superconductivity based on an antiferromagne\-ti\-cally ordered chain of magnetic atoms on the surface of a conventional
superconductor. In a \textit{weak} Zeeman field, a supercurrent in the substrate generates a staggered spin-current, which converts the preexisting topologically-unprotected
Shiba states into Majorana fermions (MFs). The two experimental knobs can be finely tuned providing a platform with enhanced functionality for applications. Remarkably, the
electronic spin-polarization of the arising edge MF wavefunctions depends solely on the parity of the number of magnetic moments, which can serve as a distinctive signature
of the MFs. We introduce the basic concepts within a minimal model and make contact with experiments by a microscopic analysis based on the Shiba states.
           
\end{abstract}


\maketitle 

The proposals for engineering Majorana fermions (MFs) in hybrid systems can be practically divided into two main categories. On one hand, we find implementations relying on
helical electronic states arising from spin-momentum locking, as for instance in: hybrid systems of topological insulators and conventional superconductors
(SCs)~\cite{FuKane2008}, superfluids~\cite{CZhang}, non-centrosymmetric SCs~\cite{NCS}, and he\-te\-ro\-structures of conventional SCs and Rashba spin-orbit coupled
semiconductors~\cite{SauSemi,AliceaSemi,Sau,Oreg}. The latter proposals stimulated experiments with encouraging, though not yet fully conclusive signatures
\cite{MFexperiments,MFexperiments2}. Other proposals \cite{Choy,Kjaergaard2012,Martin,Klinovaja,Yazdani,Nakosai2013,Klinovaja2013,Simon,Franz,Pientka,Ojanen,Das Sarma}
consider a conventional SC under the influence of a helical magnetic order \cite{Braunecker2010}, which effectively ge\-ne\-ra\-tes spin-momentum locking that can be stronger
than the intrinsic one of semiconduc\-ting nanowires. The required inhomogeneous magnetic order can be rea\-li\-zed  by pla\-cing nano-magnets \cite{Kjaergaard2012} or
mag\-netic atoms on top of a SC \cite{Martin,Yazdani,Klinovaja2013,Nakosai2013,Simon,Franz,Pientka,Ojanen,Das Sarma}. In fact, ongoing experiments involving
magnetic chains have provided the first promising MF fingerprints \cite{YazdaniExp, YazdaniExp2}. However, it has been shown that the presence of he\-li\-ci\-ty is not
indinspensable for obtai\-ning MFs \cite{KotetesClassi}, opening perspectives for new platforms.

\begin{figure}[t]
\begin{center}
\includegraphics[width=\columnwidth]{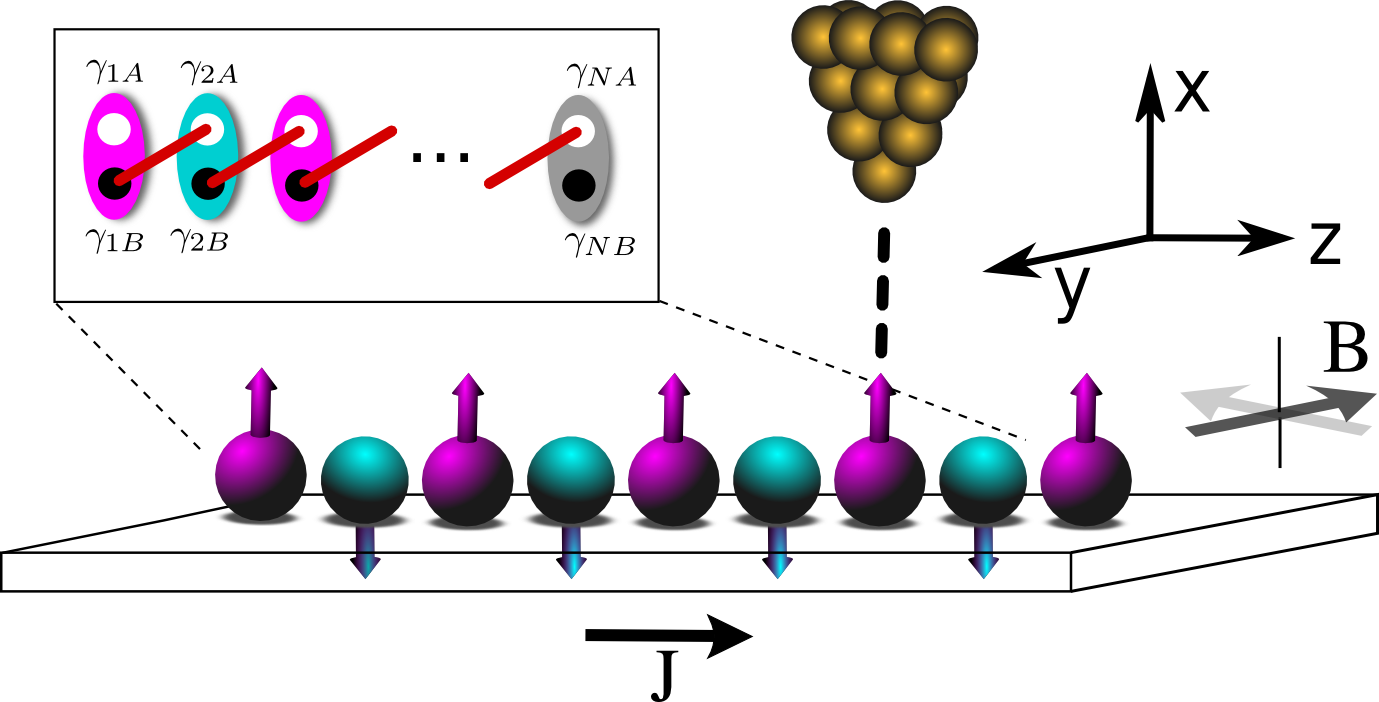}
\end{center}
\caption{Antiferromagnetic chain on top of a superconductor. The simultaneous presence of a weak in-plane Zeeman-field $B$ ($yz$-plane) and a supercurrent flow $J$
($z$-direction), convert the Shiba states into a single spin-filtered MF per edge. Inset: Mapping to Kitaev's model with unpaired MFs $\gamma_{nA,B}$.}
\label{fig_1}
\end{figure}

In this article we propose a new route towards MFs without involving helical states or helical magnetic fields. Specifically, we consider an antiferromagnetically (AFM)
ordered chain of classical spins on the surface of a conventional superconductor. The low-energy sector of this hybrid system is dominated by topologically-unprotected Shiba
states \cite{Shiba1968}, i.e. electronic states that are lo\-ca\-li\-zed at the magnetic atoms' sites with energies lying inside the superconducting gap. We demonstrate that
one can convert the Shiba states into MFs by imposing a supercurrent flow $J$ in the superconductor and applying a \textit{weak} in-plane Zeeman field $B$ (Fig.~\ref{fig_1}).
The two control fields coope\-ra\-te with the AFM order to generate a staggered spin-current, which can be viewed as an engineered time-reversal symmetry breaking
spin-momentum loc\-king. 

The functional device that we propose, offers the possibility of mani\-pu\-la\-ting the topological phase dia\-gram via the two easily controllable fields, a feature that can
facilitate the detection and braiding of MFs. Furthermore, the AFM ordering of the magnetic atoms sets stringent constraints on the spin-texture \cite{Bena} of the MF
wavefunctions, resulting in an  electronic pola\-ri\-zation of the edge states that depends on the number-parity of the magnetic moments (Fig.~\ref{fig_1}). The resulting
\textit{even-odd} effect should be experimentally detectable by spin-pola\-ri\-zed scanning tunneling microscope (STM) techiques and could help identifying the emergence of
MFs. 

It is encouraging to note that recent STM experiments \cite{Miymachi2013,Ji2008,Khajetoorians} demonstrated the existence of AFM chains on top of metallic substrates. The AFM
order is stabilized by  RKKY and the Ising nature of the magnetic moments. The latter arise when the crystal field of the substrate breaks spin-rotational symmetry and
selects an easy spin-axis. In the case of superconducting substrates, the presence of Shiba states further stabilize the AFM order \cite{Yao}.

We proceed with first examining a minimal model, which describes the arising Shiba states in the SC substrate, located at the $N$ lattice sites of the AFM chain. The chain
extends along the $z$-axis, as shown in Fig.~\ref{fig_1}. The Shiba electrons feel: \textbf{i.} an on-site super\-con\-duc\-ting gap $\Delta_n$ and \textbf{ii.} the magnetic
exchange energy scale $M$, due to the coupling to the AFM chain, which has classical magnetic moments ordered along the $x$-axis. We also consider nearest-neighbor hopping
with strength $t$. The Hamiltonian reads
\begin{eqnarray}
\label{eq::H0}
{\cal H}^0&=&-
\frac{1}{2}\sum_{n=1}^N\left[\phantom{.} (-1)^n M\Psi^\dagger_{n} \tau_z \sigma_x \Psi_{n} 
+2t \Psi^\dagger_{n}\tau_z \Psi_{n+1} \nonumber \right.\\
&&+\left. \Psi^\dagger_{n} \left(\Delta_n^\Re \tau_y \sigma_y + \Delta_n^\Im \tau_x\sigma_y\right) \Psi_{n}\right]\,.\label{eq::hamiltonian_start}
\end{eqnarray}

\noindent Here $\bm{\sigma}$ and $\bm{\tau}$ are Pauli matrices in spin and particle-hole spaces, respectively, and
$\Psi_n^\dagger=(\psi_{n\uparrow}^\dagger,\,\psi_{n\downarrow}^\dagger,\, \psi_{n\uparrow},\,\psi_{n\downarrow})$ is the spin-dependent Gor'kov-Nambu spinor. As we show later,
the particular phenomenological model can be obtained in the limit of short superconducting coherence length, from a microscopic model properly accounting for the Shiba
states (see \cite{supplementary_material}).

In contrast to a spiral magnetic order \cite{Kjaergaard2012,Martin,Yazdani,Nakosai2013,Klinovaja2013,Simon,Franz,Pientka,Ojanen,Das Sarma}, the AFM order alone is not
sufficient to ge\-ne\-ra\-te a tran\-sition to a topological superconductor (TSC). However, this can be achieved by imposing a supercurrent flow parallel to the chain and
applying a perpendicular Zeeman-field in the $yz$-plane. The supercurrent introduces a phase gradient in the super\-conducting order-parameter, {$\Delta_n=\Delta
\exp(-iJan)$}, which can be absorbed by a gauge transformation in the fermion fields, $\Psi_{n}\rightarrow \exp(-iJan\tau_z/2)\Psi_{n}$. The effect of a supercurrent has been 
considered previously either as a necessary ingredient for implementing a TSC \cite{KotetesClassi} or as an additional parameter for tuning the TSC phase diagram
\cite{supercurrents}. In the present case, it's role is crucial, since it modifies the hopping term in Eq.~\eqref{eq::hamiltonian_start}, $t\rightarrow t\cos(Ja/2)$, and more
importantly it adds a time-reversal symmetry (${\cal T}$) breaking hopping term $2it\sin(Ja/2)\Psi_{n}^\dagger\Psi_{n+1}$. Together with the Zeeman term, the  perturbations
can be written as
\begin{eqnarray}
 \label{eq::hamiltonian_perturbation}
 {\cal V}=\frac{1}{2} \sum_n\left[\Psi_n^\dagger \,\mu_BB\tau_z\sigma_z \Psi_n +2it\sin(Ja/2)\Psi_{n}^\dagger\Psi_{n+1}\right].
\end{eqnarray}

To proceed we first discuss the Hamiltonian $\mathcal{H}=\mathcal{H}_0 + \mathcal{V}$ in the limit of an infinite chain with discrete translational
invariance. In order to account for the AFM order with wave-vector $Q=\pi/a$ and lattice constant $a$ we extend the spinor in momentum
space to $\Psi_{k}^{\dagger}=(\psi_{k+Q/2,\uparrow}^\dagger,\psi_{k+Q/2\downarrow}^\dagger,\psi^\dagger_{k-Q/2\uparrow},\,\psi^\dagger_{k-Q/2\downarrow},
\psi_{-k-Q/2,\uparrow},$ $\psi_{-k-Q/2\downarrow},\psi_{-k+Q/2\uparrow},\psi_{-k+Q/2\downarrow})$ and introduce Pauli matrices $\bm{\rho}$ operating in the additional
AFM subspace. We obtain $\mathcal{H}=\frac{1}{2}\sum_{k}\Psi_{k}^{\dagger}\left[{\cal H}^0(k)+{\cal V}(k)\right]\Psi_{k}$ with 
\begin{eqnarray}
\label{eq::hamiltonian_momentum}
{\cal H}^0(k)&=&2t\cos(Ja/2)\sin(ka) \tau_z\rho_z-M \tau_z \rho_x \sigma_x-\Delta \tau_y\sigma_y\,,\nonumber\\
\mathcal{V}(k)&=&\mu_BB \tau_z\sigma_z-2t\sin(Ja/2)\cos(ka)\rho_z\,.\label{eq:Ham}
\end{eqnarray}

As a first step we consider $B=J=0$. After perfor\-ming the unitary transformation $\Psi_{k}=U\Psi_{k'}$ with
\begin{eqnarray}
U=\exp\left(i\frac{\pi}{4}\tau_z\sigma_z\right)\exp\left(i\frac{\pi}{4}\tau_y\rho_y\right)\exp\left(i\frac{\pi}{4}\sigma_x(1+\tau_z)\right)\,,\,
\end{eqnarray}
the Hamiltonian becomes diagonal in the $\rho$ and $\sigma$ spaces, yielding the {\rm BDI}-symmetry-class block Hamilto\-nian $\mathcal H_{\rho,\sigma}^0(k)=\bm
g_{\rho\sigma}^0(k) \cdot \bm \tau$~\cite{Classi,KotetesClassi} (see also \cite{supplementary_material}), where
\begin{eqnarray}
\bm g_{\rho,\sigma}^0(k)=\left(0,(\rho M-\Delta)\sigma, 2t\,\rho\sin(ka)\right)\,.\label{eq::vector0}
\end{eqnarray}

\noindent The eigenvalues of this Hamiltonian are degenerate in spin space. They are given by 
\begin{eqnarray}
%  \label{eq::spectrum}
  E_{\rho,\sigma}^0(k)=\pm\sqrt{(\rho M-\Delta)^2+ 4 t^2\sin^2(ka)}\,,
\end{eqnarray}  
  
\noindent showing a gap-closing at $\Delta=M$ for $\rho=1$. In order to discern whether this gap closing leads to a MF zero-mode, which would imply a transition to a
TSC, we make use of the specific form of Eq.~\eqref{eq::vector0} and introduce the relevant $\mathbb{Z}$ topological invariant \cite{Volovik book} defined by the
winding number
\begin{align}
\widetilde{N}_{\rho,\sigma}^0=\frac{1}{2\pi}\int_{BZ} {\rm d}k \, \left(\hat{\bm{g}}_{\rho,\sigma}^0(k)\times 
\frac{\partial \hat{\bm{g}}_{\rho,\sigma}^0(k)}{\partial k}\right)_x,
\end{align}
with $\hat{\bm {g}}_{\rho,\sigma}^0(k)=\bm{g}_{\rho\sigma}^0(k) / |\bm{g}_{\rho,\sigma}^0(k)|$. It is straightforward to show that the latter is zero, and consequently 
at this stage, with only AFM order present, there is no transition to a topologically non-trivial SC phase. 

Next we switch on the control fields $J$ and $B$. For illustration, and without loss of generality, we choose $t$ and $B$ to be small, $t,\mu_BB\ll \Delta, M$ and perform a
second order expansion based on a canonical transformation $\widetilde{{\cal H}} = \mathcal{H}^0+ i/2\left[S,\mathcal{V}\right]$
with $\left[S,\mathcal{H}^0\right]=i\mathcal{V}$ (see \cite{supplementary_material}). The perturbation modifies the energy-spectrum, but only changes in the vicinity of $k=0$
modify the topological properties. Based on this argument, we retain only the most relevant term for small $k$, characterized by the coefficient $\Lambda={2\mu_BB
t\sin(Ja/2)}/{M}$, and neglect all other second order terms in the expansion parameters $ka,\,Jat/M$ and $\mu_BB/M$. The expansion yields the mo\-di\-fied vectors
\begin{eqnarray}
\bm g_{\rho,\sigma}^{\phantom{0}}(k)=\left(0,(\rho \widetilde{M}-\Delta)\sigma+\Lambda\cos (ka), 2\widetilde{t}\,\rho\sin(ka)\right).\label{eq::vector}
\end{eqnarray}
The term proportional to $\Lambda$ lifts the spin-degeneracy in the spectrum, and we obtain
\begin{eqnarray}
\label{eq::dispersion_2}
 E_{\rho,\sigma}^{\phantom{0}}(k) = \pm \sqrt{[(\rho \widetilde{M}-\Delta)\sigma+\Lambda\cos (ka)]^2 + 4 \widetilde t\,^2\sin^2(ka)}\,.\nonumber
\end{eqnarray}

\noindent Above we introduced the renormalized va\-lues $\widetilde M=M+\left(\mu_B^2B^2+4t^2\sin^2\left(Ja/2\right)\right)/{2M}$ and $\widetilde t= t\,\cos\left(Ja/2\right)$.
The topological invariant is now given by
\begin{eqnarray}
\label{eq::winding_number}
 \widetilde{N}_{\rho,\sigma}=\frac{
{\rm sgn}\left(\rho \widetilde M-\Delta+\sigma\Lambda\right)-
{\rm sgn}\left(\rho \widetilde M-\Delta-\sigma\Lambda\right)}{2\rho\sigma}.\phantom{.}
\end{eqnarray}

\noindent We infer that the transition to the topolo\-gi\-cal non-trivial region occurs for $\widetilde M+|\Lambda|>\Delta$. We observe that due to the intrinsic magnetic
order there is no requirement for a large external field. In fact, the magnetic field is primarily required for lifting the spin-degeneracy. We also note that the combination
of Zeeman field and supercurrent flow gives rise to an effective staggered spin-current term (Eq.~\eqref{eq::Majorana_n}), which is a time-reversal violating analog of the
Rashba spin-orbit coupling in TSC nanowires \cite{Sau,Oreg}.   

\begin{figure}
\begin{center}
\includegraphics[width=\columnwidth]{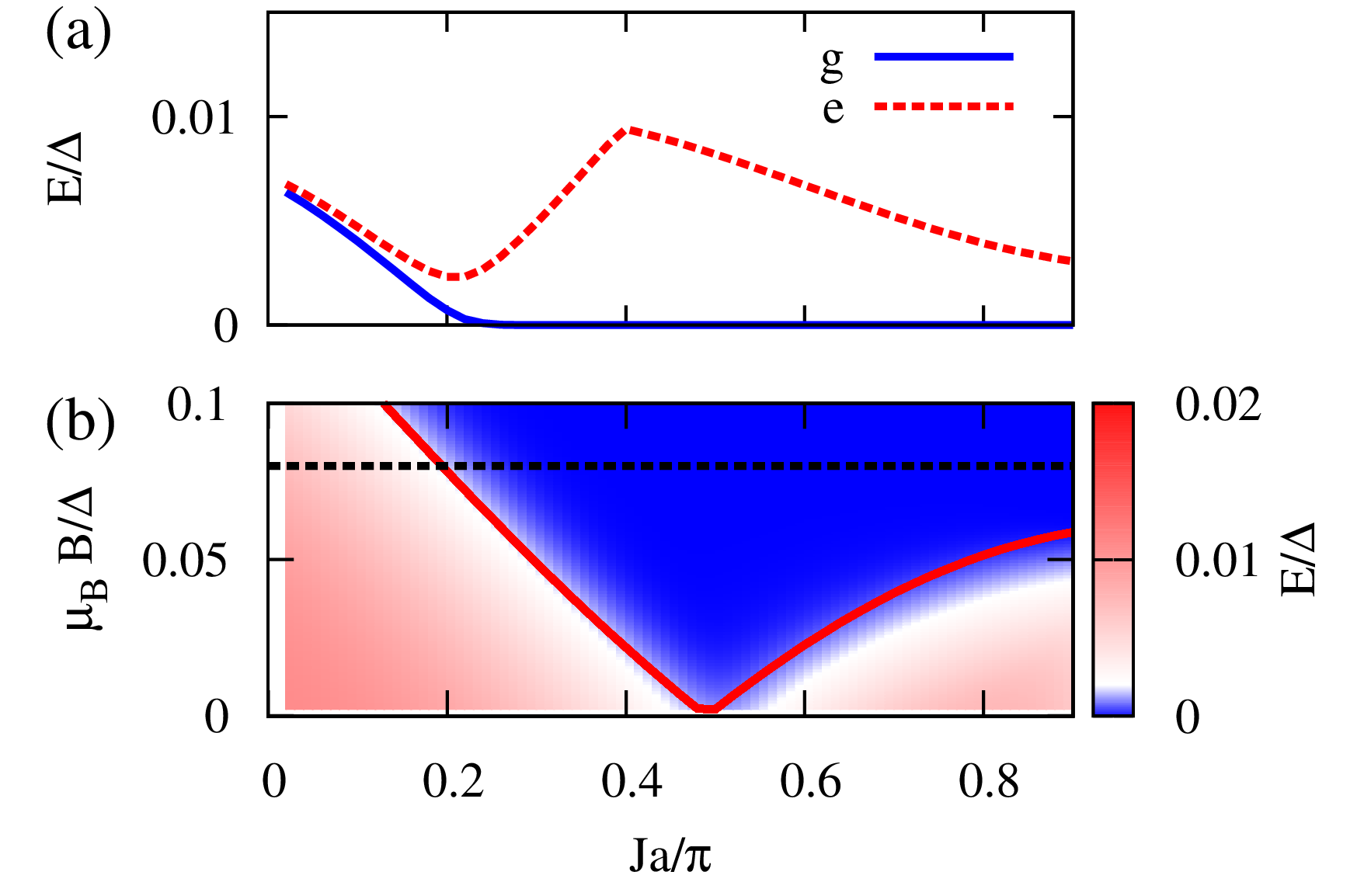}
\end{center}
\caption{\textit{Results based on the lattice version of the minimal model, Eq.~\eqref{eq::hamiltonian_momentum}:} (a) Ground-state (g) and first-excited-state (e) energies, 
as a function of $J$ for $\mu_B B/\Delta=0.08$, $M/\Delta=0.99$ and $t/\Delta=0.1$. (b) Ground-state energy in color scale depen\-ding on $J$ and $B$. The red line, $\widetilde M +
|\Lambda|=\Delta$, corresponds to the analytically derived critical value for entering the TSC phase. The value $\mu_B B/\Delta=0.08$ used in panel (a) is indicated by the
black dashed line. ($N=160$ lattice sites)}
\label{fig_2}
\end{figure}

Based on the bulk-boundary correspondence, we expect that the non-zero topological invariant obtained for suitable va\-lues of $J$ and $B$, will become manifest in the
finite-size properties of the system. To demonstrate this correspondence we show now that for the $\bm{g}$ vectors derived above, our model harbors \textit{unpaired} MFs
similarly to the situation known from Kitaev's model \cite{Kitaev2001}. To do so we first transfer to the Majorana basis
$\Psi_k^{\dag}\rightarrow\Gamma_n^{\rm T}=(\gamma_{n\uparrow},\gamma_{n\downarrow},\bar \gamma_{n\uparrow},\bar\gamma_{n\downarrow})$, where the superscript $\rm T$ denotes
transposition, and we introduced the MF operators $\gamma_{n\sigma}\equiv (\psi_{n\sigma}+\psi_{n\sigma}^{\dag})/\sqrt{2}$ and $\bar\gamma_{n\sigma}\equiv
(\psi_{n\sigma}-\psi_{n\sigma}^{\dag})/(\sqrt{2}i)$. In this basis the Hamiltonian reads
\begin{eqnarray}
  \label{eq::Majorana_n}
 \mathcal{\widetilde H} &=&\frac{1}{2} \sum_{n=1}^N \Gamma_n^T\big[\Delta \tau_x\sigma_y +(-1)^n\widetilde M \tau_y \sigma_x\big] \Gamma_n \nonumber \\
&+&\frac{1}{2}\sum_{n=1}^{N-1} \Gamma_{n}^T\big[2\widetilde t \, \tau_y -(-1)^n\Lambda\sigma_y\big]\Gamma_{n+1}\,.
\end{eqnarray}

\noindent The first line describes the on-site coupling of MFs. With the objective of unpairing them at each lattice site, we choose $\Delta=\widetilde M$, so that the
operators $\gamma_{n\uparrow}$ and $\bar\gamma_{n\downarrow}$ for odd $n$ and $\gamma_{n\downarrow}$ and $\bar\gamma_{n\uparrow}$ for even $n$ become unpaired. Within this
subspace, we choose $2\widetilde{t}=\Lambda$, and the second line becomes
\begin{eqnarray}
\label{eq::kitaev_chain}
 \mathcal{\widetilde H}_{\rm sub}
= -i\Lambda \sum_{n=1}^{N-1} \gamma_{n,B}\gamma_{n+1,A}\,.
\end{eqnarray}

\noindent In the last step we introduced the new MF operators: $\gamma_{2m-1,A}=\big(\gamma_{2m-1,\uparrow}+\bar \gamma_{2m-1,\downarrow}\big)/\sqrt{2}$
and $\gamma_{2m-1,B}=\big(\gamma_{2m-1,\uparrow}-\bar \gamma_{2m-1,\downarrow}\big)/\sqrt{2}$ as well as $\gamma_{2m,A}=\big(\gamma_{2m,\downarrow}+ 
\bar\gamma_{2m,\uparrow}\big)/\sqrt{2}$ and $\gamma_{2m,B}=\big(\gamma_{2m,\downarrow}- \bar \gamma_{2m,\uparrow}\big)/\sqrt{2}$. The transparent form of the Hamiltonian of
Eq.~\eqref{eq::kitaev_chain} displays directly the \textit{unpaired} MFs $\gamma_{1,A}$ and $\gamma_{N,B}$ at the two ends of the chain, as also illustrated in
Fig.~\ref{fig_1}. 

\begin{figure}
\begin{center}
\includegraphics[width=1\columnwidth]{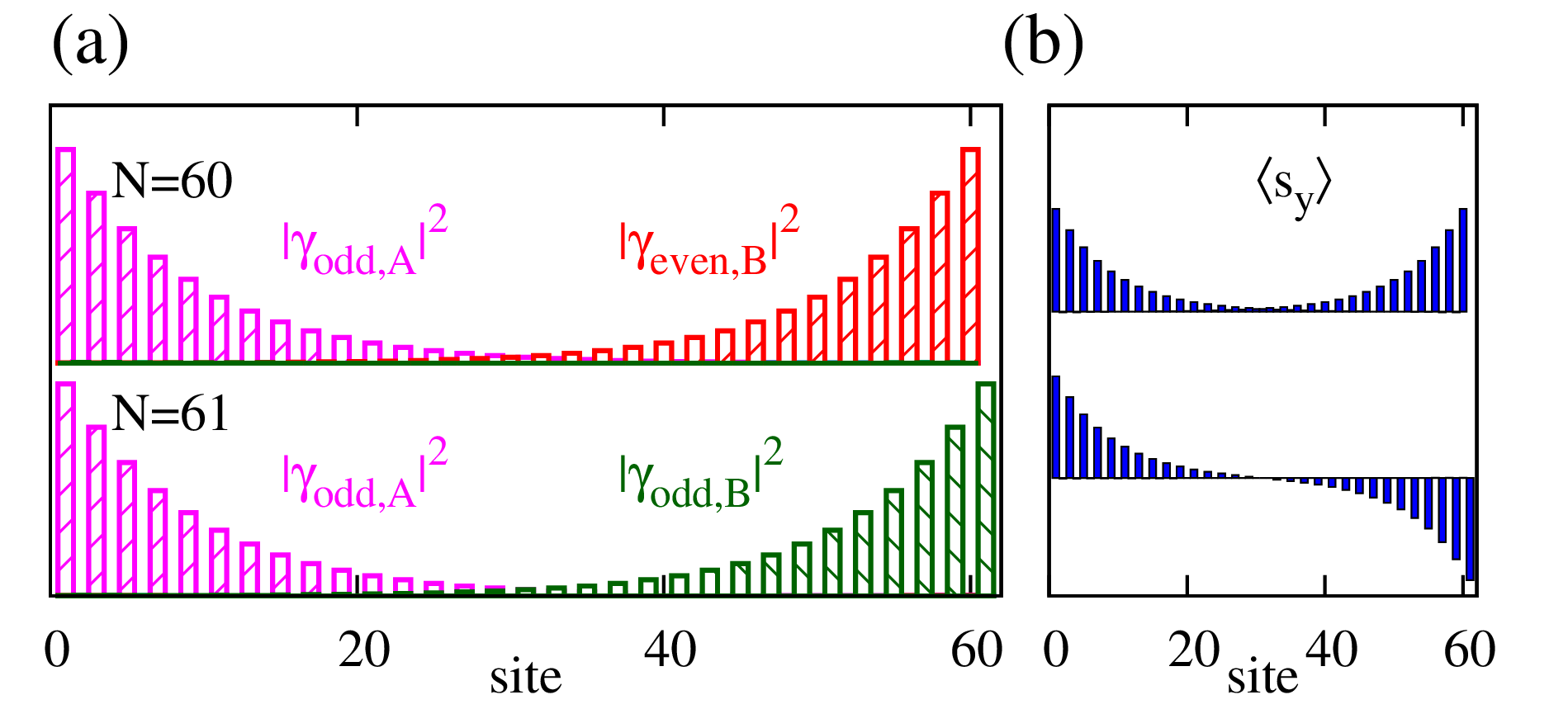}
\end{center}
\caption{\textit{Results based on the lattice version of the minimal model, Eq.~\eqref{eq::hamiltonian_momentum}:} (a) Majorana wavefunctions for an odd ($N=60$) or 
even ($N=61$) number of sites. Parameters are the same as in Fig.~\ref{fig_2}(a) with $Ja/\pi=0.4$. Depending on the parity of $N$ the right-edge Majorana wavefunction is
$\gamma_{even,B}$ ($N=60$) or $\gamma_{odd,B}$ ($N=61$). (b) Corresponding electronic spin-texture.}
\label{fig_3}
\end{figure}

The qualitative analysis based on the minimal model Hamiltonian of Eq.~\eqref{eq::hamiltonian_momentum} can be confirmed by direct numerical diagonalization. In
Fig.~\ref{fig_2} we show the lowest po\-si\-tive ener\-gies, i.e. the ground state (g) and the first excited state (e). With increasing supercurrent $J$ the system undergoes a
transition to a gapped phase accompanied by a single zero-energy MF solution per edge. This transition occurs close to the pre\-viously extracted condition $\widetilde
M+|\Lambda|=\Delta$ or equivalently $\mu_BB=|\sqrt{2M|\Delta-M|}-2t\sin(Ja/2)|$ (see red line in Fig.~\ref{fig_2}(b)). One observes that the Majorana wavefunctions are
exponentially suppressed within the bulk (see Fig.~\ref{fig_3}(a)). In addition, due to chiral symmetry they are constrained to be zero at every second site (see \cite{supplementary_material}).
Depending on the parity of $N$, the right-edge Majorana wavefunction is either $\gamma_{even,B}$ or $\gamma_{odd,B}$, i.e. it is confined to either even or odd sites. In
addition, since the electronic components of the MF wavefunctions constitute eigenstates of $\sigma_y$, the MF states have opposite electronic spin-polarization at the edges,
as plotted in Fig.~\ref{fig_3}(b).

The limit of short coherence length allowed us to transparently expose the underlying TSC mechanism and the related qualitative characteristics of the MFs. However, usually
the SC coherence length is rather long, e.g. for Pb it is $\xi_0\sim 80\,\rm nm$, to be compared to the typical spacing of the atoms $a\sim 1\,\rm nm$
\cite{Ji2008,Khajetoorians}. Below we investigate the general case by analyzing the fully \textit{microscopic} model.

The Hamiltonian of the substrate SC, with energy gap $\Delta$, is given by $\mathcal{H}_S=\sum_{\bm k} \mathcal{C}_{\bm k}^\dagger\left(\xi_{\bm
k} \tau_z-\Delta \tau_y\sigma_y\right)\mathcal{C}_{\bm k}$, where $\mathcal{C}_{\bm k}^\dagger$ $=(c_{\bm k\uparrow}^\dagger,\, c_{\bm k\downarrow}^\dagger,\, c_{-\bm
k\uparrow},\,c_{-\bm k\downarrow})$ and $c_{\bm k\sigma}^\dagger$ creates an electron with momentum $\bm k$ and spin $\sigma$, while $\xi_{\bm k}$ denotes the electronic
dispersion, leading to the quasiparticle excitation spectrum $E_{\bm k}=\sqrt{\xi_{\bm k}^2 + \Delta^2}$. Again we treat the atoms classically. They yield the
magnetic ($M$) and non-magnetic ($U$) exchange energies with the conduction electrons,  ${V}(\bm{r})=\sum_{n=1}^N
\delta(z-na)\delta(x)\delta(y) \mathcal{V}_n$ with $\mathcal{V}_n=U\tau_z -(-1)^n M \tau_z \sigma_x$.
The total Hamiltonian reads 
\begin{eqnarray}
\mathcal{H}=\mathcal{H}_S + \sum_{\bm k,\bm k'}\sum_ne^{-i(k_z-k_z')na} \mathcal{C}_{\bm k'}^\dagger \mathcal{V}_{n}\mathcal{C}_{\bm k}\,.\label{eq::micro}
\end{eqnarray}

\begin{figure}
\begin{center}
\includegraphics[width=\columnwidth]{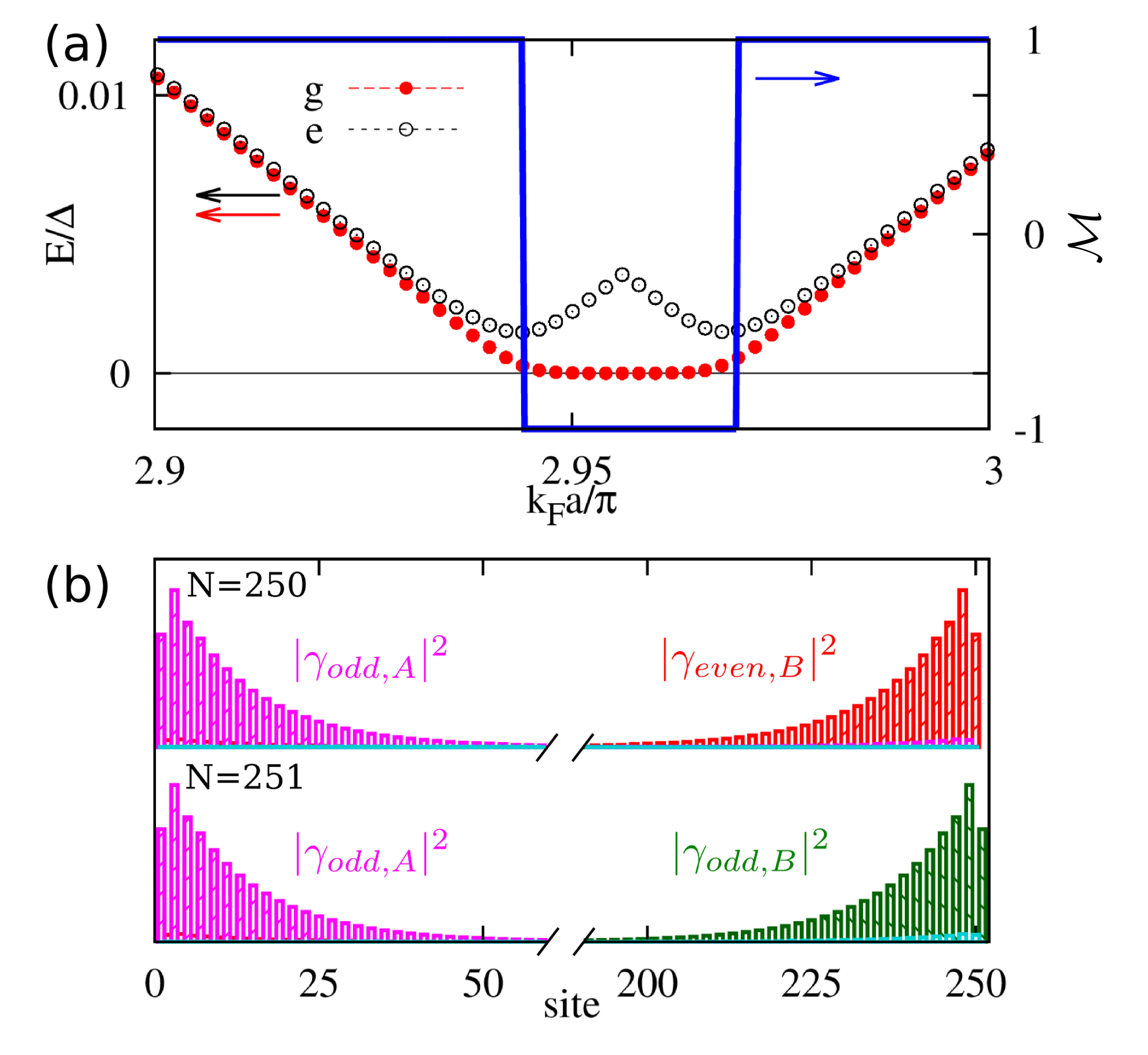}
\end{center}
\caption{\textit{Results based on the lattice version of the microscopic model, Eq.~\eqref{eq::effective_hamiltonian}:} (a) Energies of the ground-state (g-red) and 
excited-state (e-black). The corresponding topological invariant ${{\cal M}}$ is shown by the blue line (b) Majorana wavefunctions in the topolo\-gi\-cal\-ly non-trivial
regime.}
\label{fig2}
\end{figure}

\noindent We solve this problem (see \cite{supplementary_material}) in terms of the Bogo\-liu\-bov - de Gennes (BdG) equation
\cite{Flatte1997,Balatsky2006,Pientka,Yao},
\begin{align}
 \label{eq::BdG_bulk}
 \sum_{\bm k'}\bigg[\delta_{\bm{k},\bm{k}'}-G_{\bm k}(\varepsilon) \sum_{n} e^{-i(k_z-k_z')na} \mathcal{V}_n \bigg] \phi_{\bm k'} = 0,
\end{align}
where $G_{\bm k}(\varepsilon)=(\varepsilon+\Delta \tau_y\sigma_y - \xi_{\bm k} \tau_z)^{-1}$ is the Green's function of the bulk superconductor, and the spinor
$\phi_{\bm k}=(u_{\bm k,\uparrow},u_{\bm k,\downarrow},v_{\bm k,\uparrow},v_{\bm k,\downarrow})^T$ contains the spin-dependent particle and hole amplitudes $u$ and $v$.  
In the presence of a weak in-plane magnetic field and a small supercurrent $J \ll k_F$ we have to substitute $\xi_{\bm k}\rightarrow \xi_{\bm k-J{\bm{\hat z}} \tau_z/2}+
\mu_BB\sigma_z$, and 
the linearized Green's function $G_{\bm k}$ reads
\begin{eqnarray}
\label{eq::greens_function}
 G_{\bm k}(\varepsilon)
\approx\frac{\varepsilon-\Delta\tau_y\sigma_y + \xi_{\bm k} \tau_z-\mu_BB\tau_z\sigma_z + \frac{\hbar^2 J }{2m}k_{Fz}}{-E_{\bm k}^2}.
\end{eqnarray}

\noindent As we are mainly interested in the contribution of the Shiba states, which lie energetically within the energy gap, we again perform a linear expansion
in $\varepsilon/\Delta$. In order to further simplify the solution of Eq.~\eqref{eq::BdG_bulk} we trace out the continuum states, i.e. $\phi_n = \sum_{\bm k} e^{ik_z
na} \phi_{\bm k}$ and $G_n(\varepsilon)= \sum_{\bm k}e^{ik_z na} G_{\bm k}(\varepsilon)$, such that Eq.~\eqref{eq::BdG_bulk} reduces to $\sum_{s=1}^N
[\delta_{ns} - G_{n-s}(\varepsilon)\mathcal{V}_s]\phi_s=0$. This equation can be written in the form of a gene\-ra\-li\-zed eigenvalue problem $\sum_{s=1}^N {A}_{ns} \phi_s =
\frac{\varepsilon}{\Delta} \sum_{s=1}^N {B}_{ns} \phi_s$ which can be readily solved numerically. In momentum space it converts into the form of a usual Schr\"odinger-equation
${\cal {\widetilde H}}(k) \psi_{k} =\varepsilon(k)\psi_{k}$ with spinors $\psi_{k}$ related to the original ones by a transformation 
$\psi_{k}=\mathcal{L}_{k}\phi_{k}$ and {\it effective} Hamiltonian (see \cite{supplementary_material})
\begin{eqnarray}
\label{eq::effective_hamiltonian}
 &&\widetilde{\mathcal H}(k)=-\Delta \tau_y\sigma_y+ \mu_B B \tau_z\sigma_z+\sum_l \mu_l^{} \cos(2lka)\tau_z \nonumber \\ &&+\sum_l M_l^{}\cos(2lka) \tau_z \rho_x\sigma_x + 
 \sum_l t_l^{}\sin\left[(2l+1)ka\right] \tau_z\rho_z \nonumber \\
&& +\sum_l p_l {\sin(2lka)+\sum_l j_l \cos\left[(2l+1)ka\right] \rho_z}\,.
\end{eqnarray}

\noindent A comparison with Eq.~\eqref{eq:Ham} allows us to identify the terms $M_l^{},\,t_l^{}$ and $j_l^{}$ with the magnetic exchange, kinetic energy and the supercurrent,
respectively. The additional terms, $\mu_l^{}$ and $p_l^{}$, correspond to even-order-neighbor kinetic e\-ner\-gy and supercurrent contributions, respectively.
These parameters depend crucially on the relation between the spacing of the atoms, $a$, and the coherence length of the superconductor $\xi_0\sim \hbar v_F/\Delta$ (see
\cite{supplementary_material}). In the limit $a\gg \xi_0$ the microscopically derived Hamiltonian of Eq.~\eqref{eq::effective_hamiltonian} can be expanded in orders of
$\exp(-a/\xi_0)$ lea\-ding to Eq.~\eqref{eq::hamiltonian_momentum} with $t\sim \exp(-a/\xi_0)$. 

The microscopic Hamiltonian belongs to symmetry class {\rm D} and its topological properties can be investigated using the $\mathbb{Z}_2$ topological invariant ${{\cal M}}$
\cite{Kitaev2001}, defined as ${{\cal M}}\equiv{\rm sgn}\left\{{\rm Pf}[k=0]{\rm Pf}[k=\pi/2a]\right\}$. Note that the se\-cond Pfaffian is evaluated at $k=\pi/2a$,
since due to the AFM order we have to consider the folded Brillouin zone. In the parameter-space considered, only the $k=0$ component changes sign and we obtain
\begin{eqnarray}
{{\cal M}}={\rm sgn}\left[(\mu_BB)^4-2(\mu_BB)^2(A^2+h_5^2)+(A^2-h_5^2)^2\right].\label{eq::pfaffian}
\end{eqnarray}

\noindent Here we used the short-hand notation $A^2=\Delta^2+h_1^2-h_2^2$ and $h_5=\sum_l j_l $ with $h_1=\sum_l \mu_l^{} $ and $h_2=\sum_l
M_l^{}$. The system turns topo\-lo\-gically non-trivial when ${{\cal M}}$ changes sign, which occurs when $\mu_BB=|A-h_5|$. This condition is equivalent to the one
we found for the minimal model.

For the numerical analysis we choose $\xi_0/a=20\pi$, a supercurrent $J$ such that $\hbar^2Jk_{\rm F}/2m\Delta=0.5$ and a magnetic field $\mu_BB/\Delta=0.05$. The applied
field is of the order $B\sim 1\rm T$, i.e. much smaller than the critical in-plane magnetic field for Pb $\sim10{\rm T}$ \cite{Gardner2011}. As far as the supercurrent is
concerned, it corresponds to $J\sim 0.10\, J_c$ for Pb and $J\sim 0.03\, J_c$ for Nb, where $J_c\simeq1/\xi_0$ denotes the respective critical current. As shown in
Fig.~\ref{fig2}(a), ${{\cal M}}$ changes sign as a function of the atomic spacing, $a$, in a window where the system exhibits a gap between zero energy state and first excited
state. This window can be further broadened by increasing the Zeeman field or the supercurrent flowing through the substrate. In addition, for this region the ground state
wavefunctions, presented in Fig.~\ref{fig2}(b), feature the \textit{same} characteristics previously obtained within our minimal model.

In conclusion, we proposed a new mechanism for rea\-li\-zing a topological superconductor which neither involves helical states nor helical magnetic fields. Instead, the
hybrid system that we suggest is based on an array of antiferromagnetically ordered magnetic atoms deposited on the surface of a conventional superconductor, where edge
Majorana fermions controllably arise from Shiba states. Such atomic AFM chains on metallic substrates have already been fabricated and manipulated. Furthermore, STM techniques
can be used for performing spin-polarized zero-bias anomaly spectroscopy which can resolve the electronic spin texture of the Majorana wavefunctions. In fact, the edge
spin-polarization can be reversed by changing the length of the chain by a \textit{single} atom. The latter property is robust and can herald the emergence of MFs.
Furthermore, our device can be finely controlled by the combination of supercurrents and \textit{weak} magnetic fields, offering a rich test ground for experiments. The
enhanced functionality can make the setup also attractive for applications of Majorana fermions for quantum information processing. Finally, our proposal can be extended to a
setup where the AFM order is established by nano-magnets \cite{Kjaergaard2012} or to materials which already exhibit a microscopic coexistence of superconductivity and
intrinsic antiferromagnetism such as Fe-\cite{FeAs} or Ce-\cite{CeCoIn5} based superconductors.

 We thank J. Michelsen, A. Shnirman, W. Wulfhekel, A. Khajetoorians, K. Flensberg, D. Mendler, S. Nadj-Perge, A. Yazdani, M. Marthaler and C. Karlewski for
discussions. We also acknow\-ledge funding from the EU project NanoCTM (No. 234970).

\newpage
\widetext
\section*{Supplementary: Majorana fermions from Shiba states in an antiferromagnetic chain on top of a superconductor}

\subsection{Schrieffer - Wolff transformation}\label{section::appendix_canonical_trafo}

\noindent We perform a perturbative expansion of the Hamiltonian $\mathcal{H}=\mathcal{H}^0+\mathcal{V}$ of Eq.~(4), where $\mathcal
H_0=\frac{1}{2}\sum_k \Gamma^T_{-k}\mathcal{H}_k^0\Gamma_{k}$ and $\mathcal{V}=\frac{1}{2}\sum_k\Gamma^T_{-k}\mathcal{V}_k\Gamma_{k}$ with $\mathcal{H}_k^0=\Delta
\tau_x\sigma_y + M \tau_y \rho_x \sigma_x-2t\cos(Ja/2)\sin(ka) \tau_y\rho_z$ and $\mathcal{V}_k=-\mu_BB \tau_y\sigma_z - 2t\sin(Ja/2)\cos(ka)\rho_z$. To obtain the above
expressions, we first introduced $\gamma_{k\sigma}=(\psi_{k\sigma} + \psi_{-k\sigma}^\dagger)/\sqrt{2}$ and
$\bar\gamma_{k\sigma}=(\psi_{k\sigma}-\psi_{-k\sigma}^\dagger)/(\sqrt{2}i)$ and then the Majorana-spinors
$\Gamma_k^T=(\gamma_{k\uparrow},\gamma_{k\downarrow},\bar\gamma_{k\uparrow},\bar\gamma_{k\downarrow})$. Following Schrieffer and Wolff we are interested in a unitary
transformation $\mathcal{U}=e^{i\mathcal S}$ with $\mathcal S= \mathcal S^\dagger$ so that the linear order in $\mathcal{V}$ is eliminated in
$\widetilde{\mathcal{H}}=\mathcal{U}\mathcal{H}\mathcal{U}^\dagger$. The expansion can be expressed as follows $\widetilde{\mathcal{H}}=\mathcal{H} + i[\mathcal
S,\mathcal{H}]-\frac{1}{2}[\mathcal
S,[\mathcal S,\mathcal{H}]] + \cdots$. The operator $\mathcal S$ is chosen such that $[\mathcal S,\mathcal{H}^0]=i\mathcal{V}$ so that up to second order in $\mathcal{V}$ 
the Hamiltonian reads $\widetilde{\mathcal{H}}=\mathcal{H}^0 + i[\mathcal S,\mathcal{V}]/2$. Further parametrizing $\mathcal S$ by
$\mathcal S=\frac{1}{2}\sum_k\Gamma_{-k}^T\mathcal S_k\Gamma_k$ we find that
\begin{eqnarray}
 \mathcal{S}_k=\frac{t\sin(Ja/2)\cos(ka)}{M} \tau_y\rho_y\sigma_x -\frac{t^2\sin(Ja)\sin(2ka)}{2M\Delta}\tau_x\rho_x\sigma_z+\frac{\mu_BB}{2M}\rho_x\sigma_y -
\frac{\mu_BBt\cos(Ja/2)\sin(ka)}{M\Delta}\tau_z\rho_y
\end{eqnarray}

\noindent and with this that the transformed Hamiltonian is given by
\begin{eqnarray}
 \widetilde{\mathcal{H}}_k&=&\Delta \tau_x \sigma_y 
 + \bigg(M +\frac{[\mu_BB]^2+4t^2\sin^2(Ja/2)\cos^2(ka)}{2M}\bigg) \tau_y \rho_x \sigma_x 
 - 2t\cos(Ja/2)\sin(ka) \tau_y\rho_z \\&& 
 -\frac{\mu_BBt^2\sin(Ja) \sin(2ka)}{\Delta M} \tau_z \rho_x 
 - \frac{2\mu_BB t\sin(Ja/2)\cos(ka)}{M} \rho_y \sigma_y \nonumber\\&&
 + \frac{t\cos(Ja/2) \sin(ka)
\left\{(\mu_BB)^2+[{2t\sin(Ja/2)\cos(ka)]^2}\right\}}{\Delta M} \tau_x \rho_y \sigma_z\,.
\end{eqnarray}

\noindent The topological properties of the system are determined by changes at the points $k=0,\pi/(2a)$. Neglecting all terms which are linear in $k$ close to these 
points and at least second order in the expansion parameters $t,\,\mu_BB \ll M,\Delta$ we obtain the effective low energy Hamiltonian
\begin{align}
 \widetilde{\mathcal{H}}_k&\approx 
 \Delta \tau_x \sigma_y 
 +\bigg(M +\frac{[\mu_BB]^2+4t^2\sin^2(Ja/2)\cos^2(ka)}{2M}\bigg) \tau_y \rho_x \sigma_x \nonumber\\&
 -2t\cos(Ja/2)\sin(ka) \tau_y\rho_z  
 -\frac{2\mu_BB t\sin(Ja/2)\cos(ka)}{M} \rho_y \sigma_y,
\end{align}
which converts to Eq.~(11) by transferring to the lattice representation. After suitably rewriting the above Hamiltonian in the original spinor $\Psi_k$
space, we observe that it assumes a generalized time-reversal symmetry $\Theta=i\rho_y\sigma_y \cal{K}'$, i.e. $[\Theta,\widetilde{{\cal H}}_k]=0$, together with a
charge-conjugation symmetry $\Xi=\tau_x\rho_x{\cal K}'$, i.e. $\{\Xi,\widetilde{{\cal H}}_k\}=0$. Here $\cal{K}'$ defines complex conjugation, where the prime indicates 
that it does not act on $Q$. With this, also the chiral operator $\Pi=\Theta\Xi=\tau_x\rho_z\sigma_y$ anticommutes with the Hamiltonian, i.e. $\{\Pi,\widetilde{{\cal
H}}_k\}=0$. Evenmore, there exist two unitary symmetries $[\tau_y\rho_y,\widetilde{{\cal H}}_k]=[\tau_z\rho_z\sigma_x,\widetilde{{\cal H}}_k]=0$, yielding the additional
time-reversal, charge conjugation and chiral symmetries $[\tau_y\rho_y\Theta,\widetilde{{\cal H}}_k]=[\tau_z\rho_z\sigma_x\Theta,\widetilde{{\cal H}}_k]=0$,
$\{\tau_y\rho_y\Xi,\widetilde{{\cal H}}_k\}=\{\tau_z\rho_z\sigma_x\Xi,\widetilde{{\cal H}}_k\}=0$ and $\{\tau_y\rho_y\Pi,\widetilde{{\cal
H}}_k\}=\{\tau_z\rho_z\sigma_x\Pi,\widetilde{{\cal H}}_k\}=0$, respectively. Since $\Theta^2=\Xi^2=\Pi^2=1$ the Hamiltonian belongs to the symmetry class
$\oplus_{n=1}^4{\rm BDI}$ \cite{KotetesClassi}.

\subsection{Kitaev Chain}
\label{section::appendix_kitaev}
In this section we discuss the mapping to the Kitaev chain starting with the Hamiltonian in Eq.~(11), 
\begin{align}
\label{eq::kitaev_chain_matrix}
 \mathcal{\widetilde H}&=-\frac{i}{2}\sum_{n=1}^N 
 \begin{pmatrix}
  \gamma_{n\uparrow} \\ \gamma_{n\downarrow} \\ \bar{\gamma}_{n\uparrow} \\ \bar{\gamma}_{n\downarrow}
 \end{pmatrix}^T
 \begin{pmatrix}
  0 & 0 & 0 & \Delta+(-1)^n\widetilde M \\
  0 & 0 & -\Delta+(-1)^n\widetilde M & 0 \\
  0 & \Delta-(-1)^n\widetilde M & 0 & 0 \\
  -\Delta-(-1)^n\widetilde M & 0 & 0 & 0
 \end{pmatrix}
 \begin{pmatrix}
  \gamma_{n\uparrow} \\ \gamma_{n\downarrow} \\ \bar{\gamma}_{n\uparrow} \\ \bar{\gamma}_{n\downarrow}
 \end{pmatrix}\nonumber \\
 &\phantom{=}+\frac{i}{2}\sum_{n=1}^{N-1} 
  \begin{pmatrix}
  \gamma_{n\uparrow} \\ \gamma_{n\downarrow} \\ \bar{\gamma}_{n\uparrow} \\ \bar{\gamma}_{n\downarrow}
 \end{pmatrix}^T
 \begin{pmatrix}
  0 & (-1)^n\Lambda & -2\widetilde t& 0 \\
  -(-1)^n\Lambda & 0 & 0 & -2\widetilde t\\
  2\widetilde t & 0 & 0 & (-1)^n\Lambda \\
  0 & 2\widetilde t & -(-1)^n\Lambda & 0
 \end{pmatrix}
 \begin{pmatrix}
  \gamma_{n+1\uparrow} \\ \gamma_{n+1\downarrow} \\ \bar{\gamma}_{n+1\uparrow} \\ \bar{\gamma}_{n+1\downarrow}
 \end{pmatrix}.
\end{align}
Choosing $\Delta=\widetilde M$ we find that $\Delta+(-1)^n\widetilde M=0$ for odd $n$ and $\Delta-(-1)^n\widetilde M=0$ for even $n$, meaning that the on-site coupling, for
instance  between $\gamma_{n\uparrow}$ and $\bar{\gamma}_{n\downarrow}$, is cancelled at odd sites $n$. The same is true for $\gamma_{n\downarrow}$ and
$\bar{\gamma}_{n\uparrow}$ at even sites $n$. Because of that the nearest-neighbor-coupling in the second line of Eq.~\eqref{eq::kitaev_chain_matrix} can be fully expressed
within the subspace $\{\gamma_{\rm odd,\uparrow},\bar{\gamma}_{\rm odd,\downarrow},\gamma_{\rm even, \downarrow},\bar\gamma_{\rm even, \uparrow}\}$, i.e.
\begin{align}
\mathcal{\widetilde H}_{\rm sub}&=\frac{i}{2}\sum_{m}
  \begin{pmatrix}
  \gamma_{2m-1\uparrow} \\ \bar{\gamma}_{2m-1\downarrow}
 \end{pmatrix}^T
 \begin{pmatrix}
  -\Lambda & -2\widetilde t\\
  2\widetilde t & \Lambda \\
 \end{pmatrix}
 \begin{pmatrix}
  \gamma_{2m\downarrow} \\ \bar{\gamma}_{2m\uparrow} 
 \end{pmatrix}
 +\frac{i}{2}\sum_{m}
 \begin{pmatrix}
  \gamma_{2m\downarrow} \\ \bar{\gamma}_{2m\uparrow}
 \end{pmatrix}^T
 \begin{pmatrix}
  -\Lambda & -2\widetilde t\\
  2\widetilde t & \Lambda \\
 \end{pmatrix}
\begin{pmatrix}
  \gamma_{2m+1\uparrow} \\ \bar{\gamma}_{2m+1\downarrow} 
 \end{pmatrix}.
\end{align}
Especially for $\Lambda=2\widetilde t$ one finds that
\begin{align}
\mathcal{\widetilde H}_{\rm sub}&=\frac{i}{2}\Lambda\sum_{m}  
 \begin{pmatrix}
  \gamma_{2m-1\uparrow} \\ \bar{\gamma}_{2m-1\downarrow}
 \end{pmatrix}^T
 \begin{pmatrix}
  -1 & -1\\
  1 & 1 \\
 \end{pmatrix}
 \begin{pmatrix}
  \gamma_{2m\downarrow} \\ \bar{\gamma}_{2m\uparrow} 
 \end{pmatrix}
 +\frac{i}{2}\Lambda\sum_{m}
 \begin{pmatrix}
  \gamma_{2m\downarrow} \\ \bar{\gamma}_{2m\uparrow}
 \end{pmatrix}^T
 \begin{pmatrix}
  -1 & -1\\
  1 & 1 \\
 \end{pmatrix}
\begin{pmatrix}
  \gamma_{2m+1\uparrow} \\ \bar{\gamma}_{2m+1\downarrow} 
 \end{pmatrix} \\
 &=-i\Lambda\sum_m\left(\frac{\gamma_{2m-1\uparrow} - \bar \gamma_{2m-1\downarrow}}{\sqrt{2}}\frac{\gamma_{2m\downarrow} + \bar \gamma_{2m\uparrow}}{\sqrt{2}} + 
 \frac{\gamma_{2m\downarrow} - \bar\gamma_{2m\uparrow}}{\sqrt{2}}\frac{\gamma_{2m+1\uparrow} + \bar \gamma_{2m+1\downarrow}}{\sqrt{2}}\right) 
 = -i\Lambda \sum_{n=1}^{N-1}\gamma_{n,B}\gamma_{n+1,A}
\end{align}
Like in the main text we have introduced the new Majorana-operators $\gamma_{2m-1,A}=\big(\gamma_{2m-1,\uparrow}+\bar \gamma_{2m-1,\downarrow}\big)/\sqrt{2}$,
$\gamma_{2m-1,B}=\big(\gamma_{2m-1,\uparrow}-\bar \gamma_{2m-1,\downarrow}\big)/\sqrt{2}$, $\gamma_{2m,A}=\big(\gamma_{2m,\downarrow}+ \bar \gamma_{2m,\uparrow}\big)/\sqrt{2}$ and
$\gamma_{2m,B}=\big(\gamma_{2m,\downarrow}- \bar \gamma_{2m,\uparrow}\big)/\sqrt{2}$. It turns out that the electronic part $\left|\gamma_{\rm n,A/B}\right>_{{\rm el}}$ of the
latter Majorana wavefunctions are eigenstates of the spin-operator $s_y=\hbar\sigma_y/2$. We can see this by going back to the representation in terms of the original fermion
operators -- for instance $\gamma_{\rm odd,A}=(\psi_{\rm odd,\uparrow}-i \psi_{\rm odd,\downarrow}) +(\psi_{\rm odd,\uparrow}^\dagger+i\psi_{\rm odd,\downarrow}^\dagger)
= \psi_{\rm odd,\rightarrow} + \psi_{\rm odd,\rightarrow}^\dagger$. With the new fermion operators $\psi_{\rm odd,\rightleftarrows}=\psi_{\rm odd,\uparrow\downarrow}-i
\psi_{\rm odd,\downarrow\uparrow}$ and {$\left|\gamma_{A}\right>_{{\rm el}}\equiv\psi_{\rightleftarrows}^\dagger|0\rangle$} being an eigenket of $s_y=\hbar\sigma_y/2$ with
eigenvalue $\pm \hbar/2$. With this we find that 
\begin{eqnarray}
_{{\rm el}}\langle \gamma_{\rm n,A}|s_y|\gamma_{\rm n,A}\rangle_{{\rm el}}=-(-1)^n\frac{\hbar}{2}\qquad{\rm and}\qquad 
_{{\rm el}}\langle \gamma_{\rm n,B}|s_y|\gamma_{\rm n,B}\rangle_{{\rm el}}=(-1)^n\frac{\hbar}{2}\,.
\end{eqnarray}

According to the symmetry analysis in Sup.\ref{section::appendix_canonical_trafo} the Majorana wavefunctions constitute eigen-vectors of the chiral symmetry operators 
$\Pi_{\cal M}=\tau_z\rho_z\sigma_y=i\tau_z\sigma_y t_{\pi/Q}$, $\Pi_{\cal M}'=\tau_y\rho_x\sigma_y$ and $\Pi_{\cal M}''=\tau_x\sigma_z$ (expressed now in the MF basis).
{Here $t_{\pi/Q}\psi_n=\psi_{n+1}$ is the translation operator which in momentum space -- up to a phase -- corresponds to $t_{\pi/Q}=-i\rho_z$. The eigenstates of $t_{\pi/Q}$
have the property $t_{\pi/Q}\psi_n=\psi_{n+1}=\lambda \psi_n$. At the same time $t_{2\pi/Q}=t_{\pi/Q}^2$ commutes with the Hamiltonian such
that $t_{2\pi/Q}\psi_n=\psi_{n+2}=\lambda^2 \psi_n$ and therefore $\lambda=\pm 1$. This means that eigenfunctions $\psi_n^{(\pm)}$ of $t_{\pi/Q}$ are either constant or
alternating at each lattice site, i.e. $\psi_n^{(+)}=1$ or $\psi_n^{(-)}=(-1)^n$. Accordingly, the $\rho_x$ operator has eigen-functions $\psi_n=1\pm(-1)^n$, this means they
vanish every second site. Since the related Majorana wavefunction $A$ or $B$ constitutes an eigen-state of $\tau_y\sigma_y$, they must be also eigen-states of $\rho_x$ and
demonstrate this even-odd property. This can be seen for instance in Fig.~3.}

\subsection{Bogo\-liu\-bov - de Gennes equation}
\label{section::appendix_Gn}
Starting with the Green's function $G_{\bm k}(\varepsilon)$ in Eq.~(15) we retrieve the Fourier-components $G_n(\varepsilon)=\sum_{\bm k}
e^{ik_z na} G_{\bm k}(\varepsilon)$ in Eq.~(16) in terms of a quasiclassical expansion $\xi_k/\hbar v_F \ll k_F$, where $k_F$ and $v_F$ are the Fermi 
wave-vector and velocity \cite{Flatte1997}. To this end the momentum summation is converted to an integral over the linearized dispersion $\xi$ first, 
\begin{align}
 G_{n}(\varepsilon)=\sum_{\bm k} e^{i\bm k\cdot\bm r_n}\hat G_{\bm k}(\varepsilon) 
 &= \mathcal{N}_F\int_{-1}^1 \frac{du}{2} \int_{-\infty}^\infty d\xi\,
 e^{ik_F una} e^{i \frac{\xi}{\hbar v_F} u na}G_{\bm k}(\varepsilon)
\nonumber \\
&\approx \mathcal{N}_F\int_{-\infty}^{\infty} d\xi \int_{-1}^1 \frac{du}{2}\, e^{ik_Funa } e^{i\frac{\xi}{\hbar v_F}u na }  
\frac{\varepsilon+\frac{\hbar^2Jk_F \,u}{2m}  +\xi \tau_z-\Delta\tau_y \sigma_y-\mu_BB \tau_z\sigma_z}{-\xi^2-\Delta^2} 
\end{align}
Here the normal density of states of the superconductor is denoted by $\mathcal{N}_F$. The integral over $u=\cos(\theta)$ (with $\theta$ being the angle between $\bm r_n$ and
$\bm k$) gives
\begin{align}
  G_{n}(\varepsilon)&=\mathcal{N}_F\int d\xi\,
  \frac{\sin\big({k_F na +\frac{\xi}{\hbar v_F} na }\big)}{k_F na +\frac{\xi}{\hbar v_F} na}
  \frac{\varepsilon  +  \xi \tau_z - \Delta\tau_y\sigma_y-\mu_BB \tau_z\sigma_z}{-\xi^2-\Delta^2}\nonumber\\
  &-i\mathcal{N}_F\int d\xi\,\left[\frac{\sin\big({k_F na +\frac{\xi}{\hbar v_F} na }\big)-\big({k_F na +\frac{\xi}{\hbar v_F} na }\big)
  \cos\big({k_F na +\frac{\xi}{\hbar v_F} na }\big)}{\big({k_F na +\frac{\xi}{\hbar v_F} na }\big)^2}\right]
  \frac{\hbar^2Jk_F}{2m(\xi^2+\Delta^2)} .
\end{align}
Furthermore performing the quasiclassical expansion in $\xi/\hbar v_F \ll  k_F$ leads to

\begin{align}
\label{eq::greens_function_suppl}
{G_n(\varepsilon)} &\approx -{\pi \mathcal{N}_F}e^{-\frac{a|n|}{\xi_0}}\bigg[\bigg(\frac{\varepsilon}{\Delta} 
- \tau_y \sigma_y-\frac{\mu_B B}{\Delta} \tau_z \sigma_z\bigg) \frac{\sin(k_F
a|n|)}{k_F a|n|}+ \tau_z\frac{\cos(k_F a|n|)}{k_F a|n|} \nonumber \\ & +i\frac{\hbar^2 Jk_F }{2m\Delta}{\rm
sgn}(n)\,\frac{\sin(k_Fa|n|)-k_Fa|n|\,\cos(k_Fa|n|)}{(k_Fan)^2}\bigg] .
\end{align}

This expression is valid for $n\neq 0$ and has to be replaced by $G_n(\varepsilon)=-\pi \mathcal{N}_F (\varepsilon/\Delta - \tau_y\sigma_y 
- \mu_BB/\Delta \tau_z \sigma_z)$ for $n=0$. We find that the coupling between the spins crucially depends on the coherence length of the superconductor,
$\xi_0=\hbar v_F/\Delta$, which has to be
distinguished from the dispersion-energy $\xi$ used in the equations above. In Eq.~(16) it enters by the exponential factor $\exp(-|n-l|a/\xi_0)$. This
means that for $\xi_0\lesssim a$ coupling is reduced only to a few neighbors, whereas for $\xi_0 > a$ it is extended over many.\\

At this point it is convenient to write Eq.~(16) in terms of a generalized eigenvalue-problem,
\begin{align}
 \sum_{l=1}^N {A}_{nl} \phi_l &= \frac{\varepsilon}{\Delta} \sum_{l=1}^N {B}_{nl} \phi_l,
\end{align}
with the combinations
\begin{eqnarray}
 {A}_{nl}= \delta_{nl}\mathcal{V}_l +\pi\mathcal{N}_F \mathcal{V}_n\big[-(\tau_y\sigma_y+{\mu_B B}/{\Delta}\tau_z\sigma_z) a_{n-l} + \tau_z b_{n-l}
+c_{n-l}\big]\mathcal{V}_l \qquad{\rm and}\qquad
 {B}_{nl} = -\pi\mathcal{N}_F \mathcal{V}_n a_{n-l} \mathcal{V}_l, \label{eq::combinations}
\end{eqnarray}

\noindent while
\begin{align*}
a_{n-l}&={\sin(k_F a|n-l|)}e^{-a|n-l|/\xi_0}/{k_F a|n-l|}, \\
b_{n-l}&={\cos(k_F a|n-l|)}e^{-a|n-l|/\xi_0}/{k_F a|n-l|}, \\
c_{n-l}&= i\frac{\hbar^2 Jk_F }{2m\Delta}{\rm sgn}(n-l) e^{-\frac{a|n-l|}{\xi_0}}\frac{\sin(k_Fa|n-l|)-k_Fa|n-l|\,\cos(k_Fa|n-l|)}{(k_Fa|n-l|)^2}.
\end{align*}
Because the parameters $a_{n-l},\,b_{n-l}$ and $c_{n-l}$ only depend on the difference $n-l$ the eigenvalue equation (16) assumes a diagonal form in
momentum space. Mind that $\mathcal{V}_n=U\tau_z - (-1)^nM \tau_z  \sigma_x=U\tau_z + \exp(iQna)M \tau_z \sigma_x$, where $Q=\pi/a$ is the so called antiferromagnetic
wave-vector. Respectively we have to enlarge the spinor by this additional antiferromagnetic subspace in momentum space {(now we are confined along the $z$-axis)}
\begin{align}
\phi_k=
\begin{pmatrix}
u_{k+Q/2,\uparrow} & u_{k+Q/2,\downarrow} &u_{k-Q/2,\uparrow} &u_{k-Q/2,\downarrow} &v_{k+Q/2,\uparrow}&v_{k+Q/2,\downarrow}&v_{k-Q/2,\uparrow}&v_{k-Q/2,\downarrow}
\end{pmatrix}^{\rm T}\,.
\end{align}
Furthermore by introducing $\mathcal{V}=U\tau_z - M \tau_z\rho_x\sigma_x$ and the Fourier components
\begin{align}
 \alpha_k &= \sum_n e^{-ikan}(-i\rho_z)^n a_n=\alpha_k^{(e)} + \alpha_k^{(o)}\rho_z,\nonumber \\ 
 \beta_k &= \sum_n e^{-ikan}(-i\rho_z)^n b_n=\beta_k^{(e)} + \beta_k^{(o)}\rho_z,\nonumber \\ 
 \gamma_k &= -i\sum_n e^{-ikan}(-i\rho_z)^n c_n=\gamma_k^{(e)} + \gamma_k^{(o)}\rho_z, \label{eq::hamiltonian_momentum_components}
\end{align}
we obtain
\begin{align}
 \mathcal{A}_k \phi_k = \frac{\varepsilon_k}{\Delta} \mathcal{B}_k \phi_k
\end{align}
with $\mathcal{A}_k=\mathcal{V}-\pi\mathcal{N}_F\mathcal{V}\big[\alpha_k(\tau_y\sigma_y+\mu_B B/\Delta \tau_z\sigma_z) -\beta_k \tau_z - \gamma_k \big] \mathcal{V}$ and
$\mathcal{B}_k=-\pi\mathcal{N}_F\mathcal{V}\alpha_k\mathcal{V}$. As in the main text the Pauli-matrices $\rho_i$ account for the additional AFM subspace.
A Cholesky decomposition of the right-hand-site, i.e. $\mathcal{B}_k=(\mathcal{L}_k\mathcal{L}^\dagger_k)^{-1}/\Delta$, together with the transformation
$\psi_k=\mathcal{L}_k\phi_k$ leads to the effective Schr\"odinger-equation
\begin{align}
  \big[\mathcal{L}^\dagger_k \mathcal{A}_k \mathcal{L}_k\big] \psi_k=\big[-
 \Delta \tau_y\sigma_y+h^{(1)}_k\tau_z +h^{(2)}_k \tau_z \rho_x\sigma_x + h_k^{(3)} \tau_z\rho_z+ \mu_B B \tau_z\sigma_z  +h^{(4)}_k  + h_k^{(5)} \rho_z\big]\psi_k= 
  {\varepsilon_k}\psi_k.
\end{align}
The respective components of the Hamiltonian are given by
{\begin{align}
 \frac{h_k^{(1)}}{\Delta} &= \frac{U\alpha_k^{(e)}}{\pi\mathcal{N}_F(M^2-U^2)D_k}-\frac{\alpha_k^{(e)}\beta_k^{(e)}-\alpha_k^{(o)}\beta_k^{(o)}}{D_k}
  = \sum_{l=0}^\infty \mu_l \cos(2lka), \nonumber \\
\frac{h_k^{(2)}}{\Delta} &= \frac{-M}{\pi\mathcal{N}_F(M^2-U^2)\sqrt{D_k}}= \sum_{l=0}^\infty M_l \cos(2lka), \nonumber \\
\frac{h_k^{(3)}}{\Delta} &= \frac{-U\alpha_k^{(o)}}{\pi\mathcal{N}_F(M^2-U^2)D_k}-\frac{\alpha_k^{(e)}\beta_k^{(o)}-\alpha_k^{(o)}\beta_k^{(e)}}{D_k}
 = \sum_{l=0}^\infty t_l \sin\big([2l+1)ka\big], \nonumber \\ 
\frac{h_k^{(4)}}{\Delta} &= -\frac{\alpha_k^{(e)}\gamma_k^{(e)}-\alpha_k^{(o)}\gamma_k^{(o)}}{D_k}=\sum_{l=0}^\infty p_l {\sin(2lka)}, \nonumber \\
\frac{h_k^{(5)}}{\Delta} &= -\frac{\alpha_k^{(e)}\gamma_k^{(o)}-\alpha_k^{(o)}\gamma_k^{(e)}}{D_k}=\sum_{l=0}^\infty j_l {\cos\big[(2l+1)ka\big]}.
\end{align}}

\noindent with the denominator $D_k=([\alpha_k^{(e)}]^2-[\alpha_k^{(o)}]^2)$. We see that $h^{(2)}_k,\,h^{(3)}_k$ and $h^{(5)}_k$ correspond to the magnetic exchange, kinetic
energy, and the supercurrent contribution, respectively. In more detail $h^{(2)}_k$ couples even-order neighbors, while $h^{(3)}_k$ and $h^{(5)}_k$ couple odd-order neighbors.
The additional terms $h^{(1)}_k$ and $h^{(4)}_k$, correspond to even-order-neighbor kinetic energy and supercurrent contributions, respectively.\par
The system now crucially depends on the spacing, $a$, of the atoms as well, which affects the parameters $a_n$, $b_n$ and $c_n$ in Eq.~\eqref{eq::combinations}.
Note that for short coherence lengths $\xi_0\ll a$ the components \eqref{eq::hamiltonian_momentum_components} can be expanded in orders of $\exp(-a/\xi_0)$, i.e. $\alpha_k=a_0 +
2a_1\sin(ka)\rho_z$, $\beta_k=2b_1\sin(ka)\rho_z$ and $\gamma_k=2c_1\cos(ka)\rho_z$. In this limit setting $U=0$ the Hamiltonian (17) becomes equivalent to Eq.~(4) in the main
text. Within this expansion the components $h_k^{(1)},\,h_k^{(2)}$ and $h_k^{(5)}$, which enter the topological invariant in Eq.~(18), are given by
\begin{eqnarray}
 \frac{h_{k=0}^{(1)}}{\Delta} \approx 0\,,\qquad
 \frac{h_{k=0}^{(2)}}{\Delta} \approx -\frac{1}{\pi \mathcal{N}_F M} \qquad{\rm and}\qquad
 \frac{h_{k=0}^{(5)}}{\Delta} \approx \frac{\hbar^2 k_F J}{m\Delta} \exp(-a/\xi_0) \frac{\sin(k_Fa)-k_Fa\cos(k_Fa)}{(k_Fa)^2}\,.
\end{eqnarray}

\noindent 
Mind that in the main text we have used the notations $h_1=h_{k=0}^{(1)}$, $h_2=h_{k=0}^{(2)}$ and $h_5=h_{k=0}^{(5)}$.
Substituting $M'=(1/\pi\mathcal{N}_F M)$ and $J'={h_{k=0}^{(5)}}$ we obtain the condition for the topological phase-transition, i.e.
$\mu_BB=|\sqrt{(\Delta+M')(\Delta-M')}-J'|$ which in the limit $M'\sim \Delta$ is equivalent to the condition we obtained for the minimal model. \\

{Mind that generally the superconducting gap $\Delta$ has to be determined spatially resolved and in a self-consistent way as for instance done in \cite{Flatte1997}. 
There it has been shown that $\Delta$, however suppressed close to the magnetic-atom site, remains finite even in the vicinity of the quantum-phase transition $\Delta \sim M'$.
To this end we assume that the superconducting gap is finite at each atom-site and neglect its spatial variation which is a widely used approximation
\cite{Balatsky2006,Yao,Pientka}. By that, the initially complicated BdG equation (13) can be simplified by tracing out the continuum states of the superconductor such that the
problem effectively becomes one-dimensional, as we have shown in this section.}

\end{document}